\documentclass[twoside]{article}
\usepackage[latin1]{inputenc}
\usepackage{t1enc}
\usepackage{a4}
\usepackage{tabularx}
\usepackage{epsf}

\def\bref{\vspace{4pt}\noindent\hangindent=10mm}

\newcommand{\RCB}{$\rho$ CrB}

\newcommand{\degrm}{^{\circ}}
\newcommand{\mas}{{$mas$}}
\newcommand{\GHB}{Gatewood, Han, \& Black (2001)}
\newcommand{\HBG}{Han, Black, \& Gatewood (2001)}

\renewcommand\th{\thinspace} 
\newcommand\kms{\ifmmode{\rm km\th s^{-1}}\else km\th s$^{-1}$\fi} 
\newcommand\ms{\ifmmode{\rm m\th s^{-1}}\else m\th s$^{-1}$\fi} 
\newcommand\Mo{\ifmmode{M_{\odot}}\else $M_{\odot}$\fi}  
\newcommand\MJ{\ifmmode{M_{Jup}}\else $M_{Jup}$\fi}

\newcommand{\MAXLIMA}{{\footnotesize MAXLIMA}}

\begin{document}

\setcounter{figure}{0}
\setcounter{section}{0}
\setcounter{equation}{0}

\begin{center}
{\Large\bf
A Statistical Analysis of The Extrasolar Planets\\[0.2cm]
and The Low-Mass Secondaries}\\[0.7cm]

Tsevi Mazeh and Shay Zucker \\[0.17cm]
School of Physics and Astronomy, Raymond and Beverly Sackler\\  
Faculty of Exact Sciences, Tel Aviv University, Tel Aviv, Israel\\
e-mail: mazeh, shay@wise.tau.ac.il
\end{center}

\vspace{0.5cm}

\begin{abstract}
\noindent
{\it 
We show that the astrometric Hipparcos data of the stars
hosting planet candidates are not accurate enough to yield
statistically significant orbits. Therefore, the recent suggestion,
based on the analysis of the Hipparcos data, that the orbits of the
sample of planet candidates are not randomly oriented in space, is not
supported by the data. Assuming random orientation, we derive
the mass distribution of the planet candidates and shows that it is
flat in log M, up to about 10 \MJ. Furthermore, the mass
distribution of the planet candidates is well separated from the mass
distribution of the low-mass companions by the 'brown-dwarf
desert'. This indicates that we have here two distinct populations,
one which we identify as the giant planets and the other as stellar
secondaries.  We compare the period and eccentricity distributions of
the two populations and find them surprisingly similar.  The period
distributions between 10 and 1650 days are flat in log period,
indicating a scale-free formation mechanism in both populations. We
further show that the eccentricity distributions are similar --- both
have a density distribution peak at about 0.2--0.4, with some small
differences on both ends of the eccentricity range. We present a toy
model to mimic both distributions. The toy model is composed of
Gaussian radial and tangential velocity scatters added to a sample of
circular Keplerian companions. A scatter of a dissipative nature can
mimic the distribution of the eccentricity of the planets, while
scatter of a more chaotic nature could mimic the secondary
eccentricity distribution.
We found a significant paucity of massive giant planets with short orbital
periods.  The low-frequency of planets is noticeable for masses larger
than about 1 \MJ\ and periods shorter than 30 days. We point out how, in
principle, one can account for this paucity.
}
\end{abstract}

\section{Introduction}

More than fifty candidates for extrasolar planets have been announced
over the past six years (e.g., Schneider 2001).  In each case,
precise stellar radial-velocity measurements indicated the presence of
a low-mass unseen companion, with a minimum mass between 1 and about
10 Jupiter masses (\MJ). The identification of these unseen companions
as planets relied on their masses being in the planetary range.

However, the actual masses of the planet candidates are not known.
The radial-velocity data yield only $M_2 \sin\, i$, where $M_2$ is the
secondary mass and $i$ is the inclination angle of its orbital plane,
which cannot be derived from the spectroscopic data.
Nevertheless, the astronomical community considered the
planet-candidate masses as being close to their derived minimum masses
--- $M_2 \sin\, i$. This is so because at random orientation the most
probable inclination is $90\degrm$, and the expected value of
$\sin\,i$ is close to unity.

Very recently some doubt has been cast about the validity of the
random orientation assumption. \GHB\ and \HBG\ analysed the Hipparcos
astrometric data of the stars hosting planet candidates {\it together}
with the stellar precise radial-velocity measurements and derived in
some cases very low inclination angles for the orbital planes. \HBG\
found eight out of 30 systems with an inclination smaller or equal to
$0.5^\circ$, four of which they categorized as highly significant. The
probability of finding such small inclinations in a sample of orbits
that are {\it isotropically} oriented in space is extremely small,
indicating either a problematic derivation of the astrometric orbit,
or, as suggested by \HBG, some serious orientation bias in the
inclination distribution of the sample of detected planet candidates.

However, the analysis of the Hipparcos data can be misleading.  As has
been shown by Halbwachs et al.\ (2000), one can derive a small {\it
false} orbit with the size of the typical positional error of
Hipparcos, about 1 milli-arc-second (=\mas), caused by the scatter of
the individual measurements. Therefore, one should carefully evaluate
the statistical significance of any astrometric orbit of that size
derived from the Hipparcos data. In Section~2 we summarize our work
(Zucker \& Mazeh 2001a) that evaluates the significance of the
astrometric orbits by applying a permutation test to the Hipparcos 
data. Similarly to the results of Pourbaix (2001) and Pourbaix \&
Arenou (2001), we also find that the significance of all the Hipparcos
astrometric orbits of the planet candidates are less than 99\%,
including \RCB\ that attracted much attention after the publication of
\GHB\ suggestion. We therefore conclude that the Hipparcos data does
not prove the anisotropy of the orientations of the orbital planes of
the planet candidates.

After showing that the random orientation in space is still a
reasonable assumption, not confronted by any available measurement,
we present in Section 3 our work (Zucker \& Mazeh 2001b) that uses
this assumption to derive the mass distribution of the planet
candidates. This is done with \MAXLIMA, a MAXimum LIkelihood MAss
algorithm which we constructed to derive the mass
distribution. Similar to the results of Jorrisen, Mayor \& Udry
(2001), we show that the mass distribution of the planet candidates is
separated from the one of the secondary masses by the so-called
'brown-dwarf desert' (e.g., Marcy \& Butler 2000). This indicates that
we are dealing with two different classes of objects. One is the giant
planets, with masses not far from the planetary mass range, while the
other is the low-mass secondaries, with stellar mass range.

One could speculate that the separation between the two different mass
distributions indicates different formation processes.  The commonly
accepted paradigm is that planets were probably formed by coagulation of
smaller, possibly rocky, bodies, whereas stars were probably formed by
some kind of fragmentation of larger bodies. In other words, planets
were formed by small bodies that grew larger, whereas stars, binary
included, were formed by fragmentation of large bodies into smaller
objects (e.g., Lissauer 1993; Black 1995).  This could imply, for
example, that the distribution of orbital eccentricities of giant
planets and low-mass binaries would be substantially different.  All
the solar planets have nearly circular orbits, whereas binaries have
eccentric orbits (e.g., Mazeh, Mayor, \& Latham 1996). We
could also expect the periods of planets to be longer than 10 years, like
the giant planets in the solar system. Many studies of the newly
discovered planets showed that this is not the case (e.g., Marcy,
Cochran, \& Mayor 2000). Moreover, following Heacox (1999) who based
his analysis upon only 15 binaries and a handful of planet candidates,
we show in Section 4 that within some reasonable restrictions, the
eccentricity and period distributions of the two samples are
surprisingly similar.  Similar results have been obtained by
Stepinski \& Black (2001a,b,c).  In Section 5 we consider a toy model
that can generate the eccentricity distribution of both populations.

\section{The Significance of the Astrometric Orbits}

In this section we present our work (Zucker \& Mazeh 2001a) where we
evaluate for each of the extrasolar planets the statistical
significance of its astrometric orbit, derived from the Hipparcos data
{\it together} with its radial-velocity measurements. We first derived
the best-fit orbit by assuming that the spectroscopic and astrometric
solutions have in common the following elements: the period, $P$, the
time of periastron passage, the eccentricity, $e$, and the longitude
of the periastron.  In addition, the spectroscopic elements include
the radial-velocity amplitude, $K$, and the center-of-mass radial
velocity.  We have three additional astrometric elements --- the
angular semi-major axis of the photocenter, $a_0$, the inclination,
$i$, and the longitude of the nodes. In addition, the astrometric
solution includes the regular astrometric parameters --- the parallax,
the position and the proper motion.

In most cases the elements are not all independent. From the
spectroscopic elements we can derive the projected semi-major axis of
the primary orbit. This element, together with the inclination $i$ and
the parallax, yields the angular semi-major axis of the primary,
$a_1$.  Assuming the secondary contribution to the total light of the
system is negligible, this is equal to the observed $a_0$.

To find the statistical significance of the derived astrometric orbit
in each case we applied a permutation test (e.g., Good 1994) to the
Hipparcos data. For each star we generated simulated permuted
astrometric data and analyzed them either together with the actual
individual radial velocities of that star, or by imposing the
published spectroscopic elements. Details of the analysis are given
by Zucker \& Mazeh (2001a).

The distribution of the falsely detected semi-major axes
indicated the range of possible false detections. For example,
$a_{99}$ --- the 99-th percentile, denotes the semi-major axis size for
which 99\% of the simulations yielded smaller values. Consequently, an
astrometric orbit is detected with a significance of 99\% if and only
if the actually derived semi-major axis, $a_{derived}$, is larger than
$a_{99}$.

As an illustration, Figure~1A shows the histogram of the semi-major
axis derived by random permutations of the Hipparcos data of
HD~209458. This star's inclination is known to be close to $90 \degrm$
through the combination of radial velocity and transit measurements
(Charbonneau et al.\ 2000; Mazeh et al.\ 2000; Henry et al.\ 2000; Brown
et al.\ 2001).  The Hipparcos derived semi-major axis, $a_{derived}$,
is 1.76 \mas, which is marked in the figure by an arrow. One can
clearly see that many random permutations led to larger semi-major
axes, a fact that renders this derived value insignificant. The
derived value is obviously false since the known inclination implies a
value of less than a micro-arc-second.

In Figure 1B we show an opposite case, HD~164427, where the derived
astrometric orbit is quite significant. Note that $a_{derived}$ is
relatively large --- 3.11 \mas, which made the significant detection
possible. However, this is not a planet-candidate case. The minimum
mass suggests this secondary is a brown-dwarf candidate, whereas the
astrometric orbit shows the secondary mass is in the stellar regime.

\begin{figure}[htb]
\centerline{\epsfxsize=100mm\epsfbox{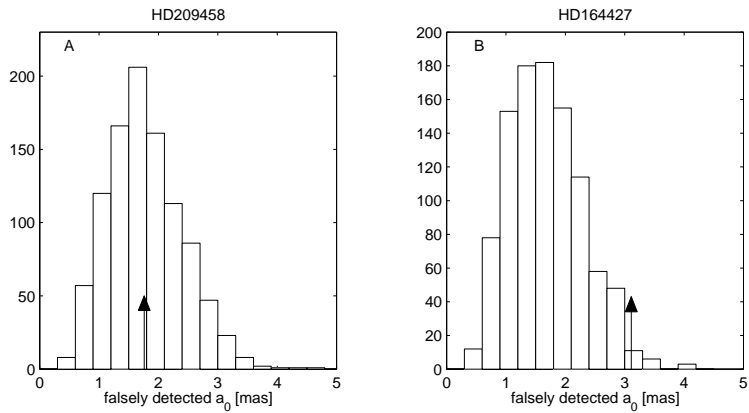}}
\caption{Histograms of the size of the falsely derived semi-major
axes in the simulated permuted data of HD~209458 (A) and HD~164427 (B). 
The size of the actually detected axes are marked by an arrow. }
\end{figure}

As of March 2001, the Encyclopedia of extrasolar planets included 49
planet candidates with minimum masses smaller than 13 \MJ. We (Zucker
\& Mazeh 2001a) analyzed all but two of the planet candidates.  One
star had no Hipparcos data, and the other star is known to have two
companions. Figure~2 presents our results by depicting $a_{derived}$
versus $a_{99}$. The figure indeed shows that all points fall to the
right of
the line $a_{derived} = a_{99}$. This means that all our derived
astrometric motions are not significant in the 99\% level.  This
includes the planets of $\upsilon$~And and HD~10697 whose derived
orbits were previously published by us (Mazeh et al.\ 1999; Zucker \&
Mazeh 2000), but the new analysis renders their orbits less
significant.

\begin{figure}[htb]
\centerline{\epsfxsize=100mm\epsfbox{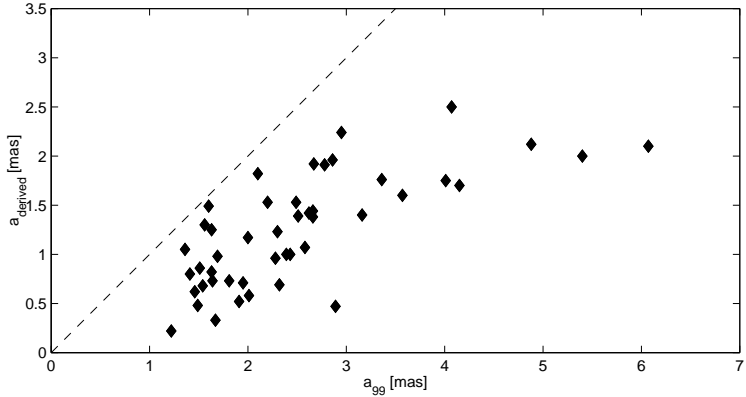}}
\caption{The derived semi-major axes
  of the planet candidates as a
  function of the 99-th percentile of the falsely derived
  semi-major axes. The dashed line represents the line $a_{derived} =
  a_{99}$.}
\end{figure}

Note, however, that this does not mean that the orbits derived are all
false.  Figure~2 shows that some of the systems are close to the
border line, indicating that the orbits of these systems were detected
with significance close to 99\%. The systems with significance higher
than 90\% are listed in Table~1. Here we list the Hipparcos number and
the stellar name, the confidence level of the derived astrometric
orbit, the derived semi-major axis, its uncertainty and the derived
inclination; the derived secondary mass together with its 1$\sigma$
range.

\begin{table}[htb]
{\scriptsize
\begin{tabular}{lllllllc}
HIP    & Name       &Signif-&$a_{derived}$&$\sigma_a$&$i_{derived}$&$M_{derived}$&Mass Range \\
number &            &icance            &(\mas) &(\mas) & (deg) &(\Mo)&(1$\sigma$) \\
\hline
5054   & HD 6434    & 0.96 & 1.34 & 0.67 & -0.08 & 0.45  & (0.20,0.77)   \\
43177  & HD 75289   & 0.90 & 1.05 & 0.52 & 0.03  & 1.13  & (0.45,2.19)   \\
78459  & $\rho$ CrB & 0.98 & 1.49 & 0.46 & 0.54  & 0.12  & (0.086,0.17)  \\
90485  & HD 169830  & 0.92 & 1.25 & 0.64 & 2.1   & 0.081 & (0.039,0.124) \\
94645  & HD 179949  & 0.90 & 1.92 & 0.68 & 0.034 & 3.4   & (1.57,6.49)   \\
98714  & HD 190228  & 0.95 & 1.82 & 0.77 & 4.5   & 0.064 & (0.037,0.093) \\
100970 & HD 195019  & 0.92 & 2.24 & 0.78 & 0.32  & 0.92  & (0.51,1.47) \\
\end{tabular}}
\caption{Derived planet-candidate orbits
with confidence level higher than 90\%.}
\end{table}

To summarize, the combination of the Hipparcos data together with the
radial-velocity measurements did not yield any astrometric orbit with
significance higher than 99\%. Apparently, the Hipparcos precision is not good
enough to detect a 1 \mas\ orbit, even with the combination of the
radial-velocity measurements. The analysis shows that the data are
consistent with no astrometric detection at all, although one or two
true astrometric orbits, which imply low inclinations, are still
possible. However, such a finding would {\it not} prove that the orbits of
the sample of planet candidates are not randomly oriented in space.

\section{The Mass Distribution of the Extrasolar Planets}

Assuming the orbits of the detected planet candidates are randomly
oriented in space we can now proceed to derive their mass
distribution. To do that we have to account for the unknown orbital
inclination and for the fact that stars with too small radial-velocity
amplitudes could not have been detected as radial-velocity
variables. Therefore, planets with masses too small, orbital periods
too large, or inclination angles too small were not detected.

Numerous studies accounted for the effect of the unknown inclination
of spectroscopic binaries (e.g., Mazeh \& Goldberg 1992; Heacox 1995;
Goldberg 2000), assuming random orientation in space. Heacox (1995)
calculated first the minimum-mass distribution and then used its
relation to the actual mass distribution to derive the latter. This
calculation amplified the noise in the observed data, and necessitated
the use of quite heavy smoothing of the observed data.  Mazeh \&
Goldberg (1992) introduced an iterative algorithm whose solution
depended, in principle, on the initial guess.

Very recently Jorissen, Mayor, \& Udry (2001a) studied the planet
distribution by considering only the effect of the unknown
inclination.  Like Heacox (1995), Jorrisen, Mayor, \& Udry derived
first the distribution of the minimum masses and then applied two
alternative algorithms to invert it to the distribution of planet
masses. One algorithm was a formal solution of an Abel integral
equation and the other was the Richardson-Lucy algorithm (e.g., Heacox
1995). The first algorithm necessitated some degree of data smoothing
and the second one required a series of iterations.  The results of
the first algorithm depended on the degree of smoothing applied, and
those of the second one on the number of iterations performed. In
addition, Jorissen, Mayor, \& Udry (2001) did not apply any correction
to the observational selection effect.

We (Zucker \& Mazeh 2001b) followed Tokovinin (1991, 1992) and
constructed a maximum likelihood algorithm --- MAXimum LIkelihood
MAss, to derive an histogram of the mass distribution of the
extrasolar planets.  \MAXLIMA\ derives the histogram directly by
solving a set of numerically stable linear equations. It does not
require any smoothing of the data, except for the bin size of the
histogram, nor any iterative procedure. \MAXLIMA\ also offers a
natural way to correct for the undetected planets. This is done by
considering each of the detected systems as representing more than one
system with the same $M_2 \sin\, i$, depending mainly on the period
distribution.  The details of the algorithm are given in Zucker \&
Mazeh (2001b).

To apply \MAXLIMA\ to the current known sample of extrasolar planets
we (Zucker \& Mazeh 2001b) considered all known planets and brown
dwarfs orbiting G- or K-star primaries as of April 2001. To acquire
some degree of completeness to our sample we have decided to exclude
planets with periods longer than 1500 days and with radial-velocity
amplitudes smaller than 40 m/s. The values of these two parameters
determine the correction of \MAXLIMA\ for the selection effect, for
which we assumed a period distribution which is flat in $\log P$. This
choice of parameters also implies that our analysis applies only to
planets with periods shorter than 1500 days. We further assumed that
the primary mass is 1\Mo\ for all systems.

The results of \MAXLIMA\ are presented in the lower panel of Figure~3
on a logarithmic mass scale. The value of each bin is proportional to
the estimated number of planets found in the corresponding range of
masses in the known sample of planet candidates, after correcting for
the undetected systems. To estimate the uncertainty of each bin we ran
5000 Monte Carlo simulations and found the r.m.s.\ of the derived
values of each bin. Therefore, the errors plotted in the figure
represent only the statistical noise of the sample. Obviously, any
deviation from the assumptions of our model for the selection effect
induces further errors into the histogram, the assumed period
distribution in particular. This is specially true for the first bin,
where the actual number of systems is small and the correction factor
large.

\begin{figure}[htb]
\centerline{\epsfxsize=100mm\epsfbox{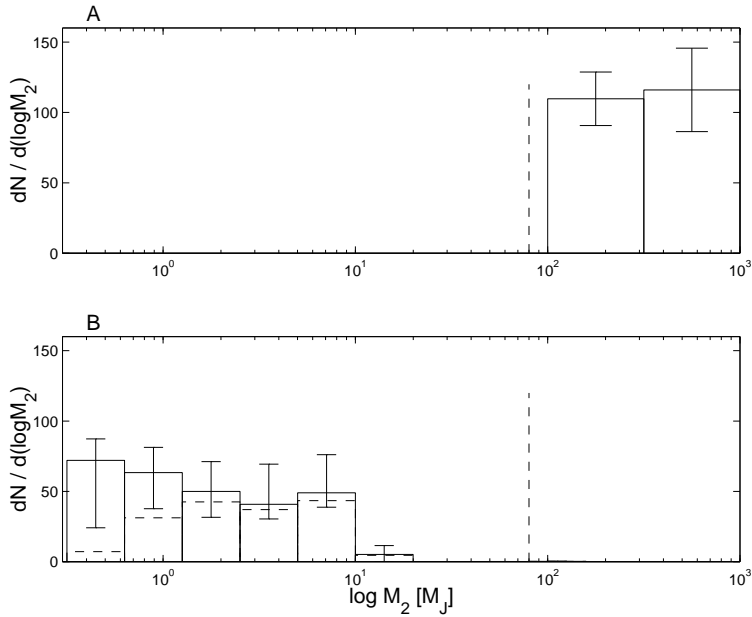}}
\caption{The mass distributions of the planets (lower panel) and the 
stellar companions (upper panel). The horizontal dashed lines
represent the mass distribution without the correction for the
selection effect. The vertical dashed line marks the
stellar--sub-stellar border line.}
\end{figure}

To compare the mass distribution of the planet candidates with that of
the stellar secondaries we plot (Zucker \& Mazeh 2001b) the latter on
the same scale in an adjacent panel of Figure~3. We plot here only two
bins, with masses between 100 and 1000 \MJ, using subsamples of
binaries found by the Center for Astrophysics (=CfA) radial-velocity
search for spectroscopic binaries (Latham 1985) in the Carney \&
Latham (1987) sample of the high-proper-motion stars (Latham et al.\
2001; Goldberg et al.\ 2001).

Note that the upper panel does {\it not} have any estimate of the
values of the bins with masses smaller than 100 \MJ. This is so
because the CfA search does not have the sensitivity to detect
secondaries in that range. On the other hand, the lower panel does
include information on the bins below 100 \MJ. This panel presents the
results of the high-precision radial-velocity searches, and these
searches could easily detect stars with secondaries in the range of,
say, 20--100 \MJ. The lower panel shows that the frequency of secondaries
in this range of masses is close to zero. 

The relative scaling of the planets and the stellar companions is not
well known (see Zucker \& Mazeh 2001b for a detailed discussion).
Nevertheless the comparison is illuminating. It suggests that we have
here two distinct populations, separated by a 'gap' of about one
decade of masses, in the range between 10 and 100 \MJ.  We will assume
that the two populations are the giant planets, at the low-mass side
of Figure~3, and the stellar companions at the high-mass end of the
figure.  The present analysis is not able to tell whether the gap 
extends up to 60, 80 or 100 \MJ.

The gap between the two populations was already noticed by many
previous studies (Basri \& Marcy 1997; Mayor, Queloz, \& Udry 1998;
Mayor, Udry, \& Queloz 1998; Marcy \& Butler 1998). Those papers binned
the mass distribution linearly.  Here we follow our previous work
(Mazeh, Goldberg \& Latham\ 1998; Mazeh 1999a,b; Mazeh \& Zucker 2001)
and use a
logarithmic scale to study the mass distribution, because of the large
range of masses, 0.5--1000 \MJ, involved.  The gap or the
brown-dwarf desert is consistent also with the finding of
Halbwachs et al.\ (2000), who used Hipparcos data and found that many
of the known brown-dwarf candidates are actually stellar companions.

The distribution we derived in Figure~3 suggests that
the planet mass distribution is almost flat in $\log M$ over five bins
--- from 0.3 to 10 \MJ. Actually, the figure suggests a possible
slight rise of the distribution toward smaller masses. At the
high-mass end of the planet distribution the mass distribution
dramatically drops off at 10 \MJ, with a small high-end tail in the
next bin. Although the results are still consistent with zero, we feel
that the small value beyond 10 \MJ\ might be real. The dramatic drop
at 10 \MJ\ and the small high-mass tail agree with the findings of
Jorissen, Mayor, \& Udry (2001), despite the differences in the algorithm used
to derive the distribution, and the logarithmic scale we use for the
distribution.

\section{Eccentricity and Period Distribution of the Two Populations}

Having established the difference between the mass distribution of the
giant planets and that of the low-mass secondaries in spectroscopic
binaries, we turn now to compare the period and eccentricity
distributions of the two populations.  For the latter we use (Mazeh \&
Zucker 2001) the results of a very large radial-velocity study of the
Carney \& Latham (1987) high-proper-motion sample, which yielded 
about 200
spectroscopic binaries (Latham et al.\ 2001; Goldberg et al.\
2001). Goldberg (2000) separated statistically between the binaries of
the Galactic halo and those coming from the disk. We consider in this
section only the 59 single-lined spectroscopic binaries (=SB1s) of the
Galactic disk.  For the giant planet sample we use again the sample of
66 planet candidates listed in Schneider (2001) as of April 2001.

Figure 4A shows the cumulative period distribution of the two samples.
The figure suggests similar general trend, except in the two ends of
the distributions. We therefore plotted in Figure 4B the two
distributions only in the range between 10 and 1650 days. The
similarity is astounding, since the two distributions are
identical. Both are consistent with a straight line, which implies a
flat distribution in log P.

\begin{figure}[htb]
\centerline{\epsfxsize=118mm\epsfbox{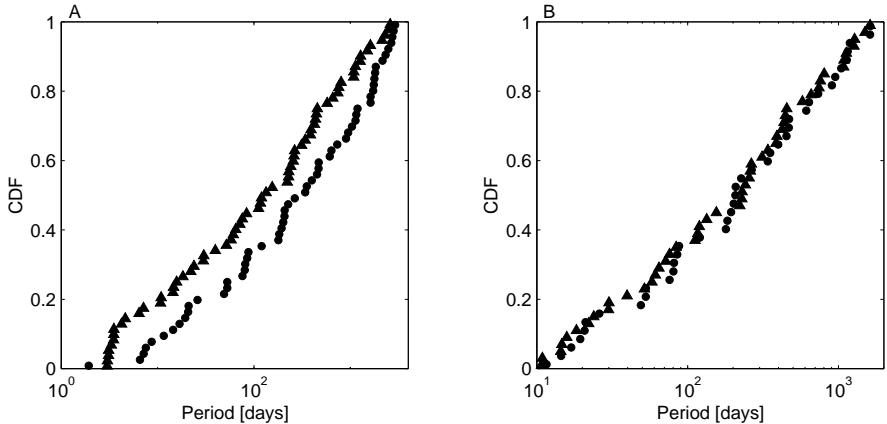}}
\caption{The period cumulative distribution for the planet candidates
  (triangles) and
the Galactic disk SB1s (circles). A. All the stars in the samples. 
B. After restricting the samples to $10<P<1560\,{\rm d}$.}
\end{figure}

We speculate that at the short period range, below 10 days, some
dynamical interaction changed the distribution of either one or both
distributions. Such an interaction could also change the eccentricity
distribution of the orbits. In order not to be distracted by this
possible interaction when we consider the eccentricity distribution, we
choose to consider only the eccentricities of the orbits with periods
between 10 and 1650 days.  The cumulative distributions are plotted in
Figure 5A. We again see a similar trend in both distributions, except
in both ends of the range [0,1]. To illuminate the difference we
plotted the density distribution in Figure 5B.  We derived the
distribution by convolving the actual data points with a Gaussian
kernel with a width of 0.08. It is clear that both distributions peak
at about 0.2--0.4. However, the distribution of the spectroscopic
binaries drops sharply toward zero, whereas the planet distribution does
not. The eccentricity distribution of the binaries displays a tentative
'shoulder' at the large eccentricities, whereas that of the
planets displays such a possible shoulder at the small eccentricities.

\begin{figure}[htb]
\centerline{\epsfxsize=118mm\epsfbox{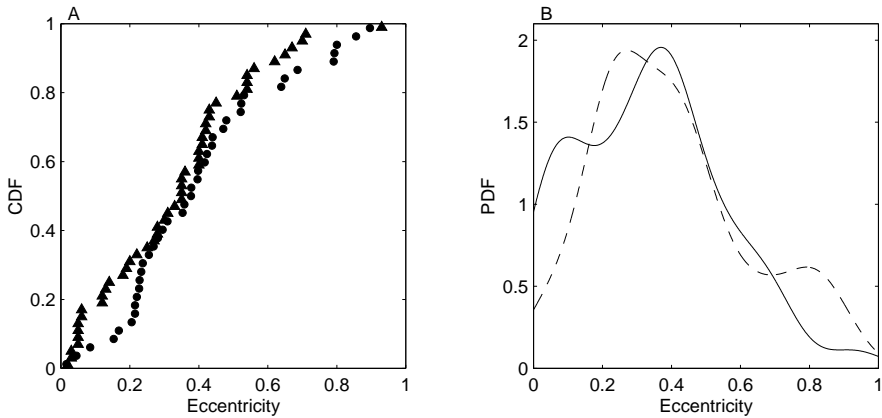}}
\caption{A. The eccentricity cumulative distribution for the planet 
candidates (triangles) and
the Galactic disk SB1s (circles), restricted to $10<P<1560\,{\rm d}$. 
B. Estimated probability 
density function of the same samples, using a 0.08-wide kernel. The continuous line represents the planets and the dashed line represents the SB1s.}
\end{figure}

Any paradigm that assumes the two populations were formed differently
has to explain why their eccentricity as well as period distributions
are so much alike. Although we do not try to explain any of the two
similarities, we suggest in the next section a toy model that can
generate the two eccentricity distributions.

\section{A Toy Model to Generate the Eccentricity Distributions of the
Two Samples}

Consider a sample of low-mass companions that orbit their parent stars
in circular Keplerian orbits. For simplicity let us choose the units
such that the orbital radii of all orbits are of length unity, and so
are their orbital tangential velocities. Now let us introduce 
a Gaussian scatter to the velocities of the companions of the
sample, with two independent components. One component is tangential
and the other is radial. The tangential component changes the moduli of
the velocities, while the radial one changes mainly their directions.

The new scatter determined the new velocity {\it distribution}.
Denote the center of the distribution by $v_0$ and its r.m.s.\ by
$\sigma_v$. Suppose that the velocity angles are distributed around
$90\degrm$, with r.m.s of $\sigma_{\theta}$. Note that the
distribution has three parameters, $v_0$, $\sigma_v$ and
$\sigma_{\theta}$.

We can now calculate the eccentricity distribution of the sample, and
see if such a simple-minded toy model can mimic the observed
distributions of the giant planets and the low-mass
companions. Figure~6 compares the two. We found that we can approximate
the giant planet distribution with $v_0=0.82$ and $\sigma_v =0.13$,
while $\sigma_{\theta}=0$, whereas the low-mass
stellar companions necessitated $\sigma_{\theta}=25\degrm$,
$\sigma_v =0.05$ and $v_0=1.08$.

\begin{figure}[htb]
\centerline{\epsfxsize=118mm\epsfbox{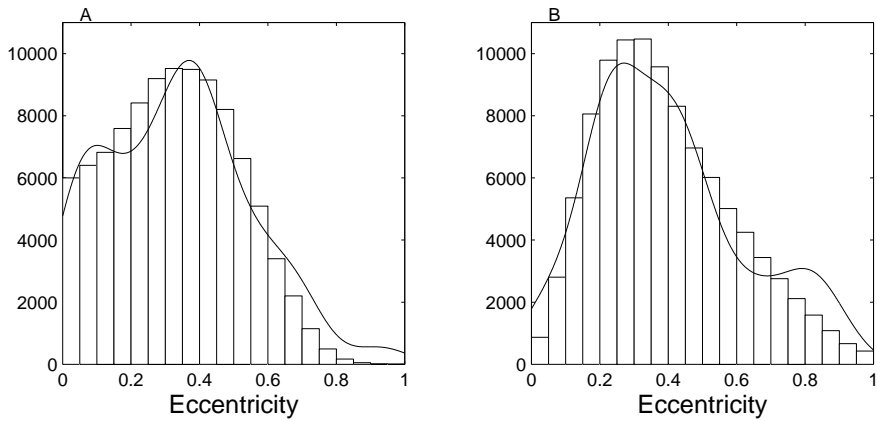}}
\caption{The simulated eccentricities histograms, together with the empirically 
estimated distributions of the planet candidates (A) and the stellar
companions (B).}
\end{figure}

The fact that we succeeded to mimic the two actual distributions is
not surprising. As the old statistical saying goes: ``You can fit an
elephant with any model with two parameters, and you can make him
dance with three''. However, the specific values of the parameters
found are somewhat intriguing. Suppose that both populations started
with Keplerian {\it circular} orbits, and two mechanisms introduced
the scatter into the two populations. Suppose the nature of the
mechanism that operated on the planet population was dissipative, like
the dissipation generated by an interaction of a planet with a swarm
of small particles in a disk. Such a mechanism could {\it decrease}
the velocity without changing its direction. This would result with a
null $\sigma_{\theta}$ and $v_0$ less than unity, the difference being
of the same order of $\sigma_v$. On the other hand, the spectroscopic
binaries could be subject to a more chaotic, eruptive disturbing
mechanism, like the gravitational interaction with a few large
bodies. In such a process one could expect a spread of the velocity
directions and moduli, without significantly changing $v_0$.  This
simple-minded picture is consistent with our findings.

We should emphasize that the aforementioned discussion is not meant to
explain how the eccentricities were formed, nor why the two
distributions are similar with some definite small differences. The
model might only serve as a starting point for any theoretical study
to account for the observed distributions.

\section{The Paucity of Short-Period Massive Planets}

In Section 3 \& 4 we have discussed the distributions of masses,
periods and eccentricities of the extrasolar planets. In this section
we move to examine one aspect of the inter dependence of these
variables. To explore this possible dependence we performed a
Principal Component Analysis (e.g., Kendall \& Stuart 1958), which 
immediately pointed out
to the significant correlation between the (minimum) masses and
periods of the extrasolar planets. This is depicted in Figure~7, in
which we plotted the period as a function of the (minimum) masses of
the known planets, as of April 2001. We choose to plot the two axes
with logarithmic scales, because the frequency of planets is flat in
log M and log P, as has been shown in previous sections.

\begin{figure}[htb]
\centerline{\epsfxsize=100mm\epsfbox{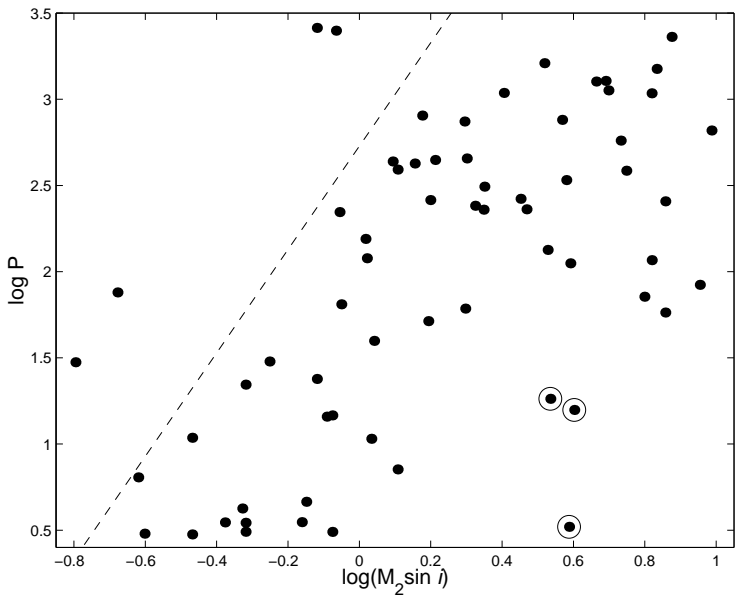}}
\caption{The logarithm of the period vs.\ the logarithm of the mass of 
the planet candidates. The dashed line represents a detection limit of
25 \ms\ radial velocity amplitude. The three circled points correspond
to the stars HD~195019, Gls~86 and $\tau$~Boo (see text).}
\end{figure}

Most of the correlation between the periods and masses of the
extrasolar planets could be accounted for by a selection effect, that
prevents planets which are not massive enough from being discovered if
their periods are too long.  Such systems have radial-velocity
amplitude, $K$, which is too small to be detected by the present
planet-search projects.  This is easily seen in the
small-mass--long-period corner of the diagram, bounded by the $K=25\,
\ms$ line. There are only four planets above this line. However, a
close examination of Figure~7 reveals an additional feature --- a
significant paucity of planets at the opposite,
large-mass--short-period corner of the diagram. Only three planets
appear at that corner, all marked by a circle. This is certainly not a
selection effect, because planets at that part of the diagram have the
largest radial-velocity amplitude, and therefore are the easiest to
detect.

It is not clear yet what is the shape of the area in which we find
low frequency of planets. That corner might have a rectangular
shape bordered by $\log(M_2\sin i)=0.2$ and $\log P= 1.5$, or could be
of a wedge shape, bordered by the line that goes from 
$\Large(\log(M_2\sin i),\log P\Large) = (0,0.5)$ to $(1,1.5)$.
 
The three planets that we find in the small-mass--long-period corner
are Gls~86, HD~195019 and $\tau$~Boo. Interestingly enough, all three
systems are wide binaries. Els et al. (2001) discovered very recently
that the star Gls~86 has a brown-dwarf companion at about 20 AU
projected separation.  Pourbaix \& Arenou (2001) pointed out that
HD~195019 is a known visual binary with a companion fainter by about 3
mag., observed at a separation of 3.5 arc-sec in 1988 (Mason et al.\
2001).  The angular separation of HD~195019 (=WDS 20283+1846)
translates to 130 AU projected separation for a parallax of 27 $mas$
(ESA 1997).  The third star, $\tau$~Boo, is also a known visual binary
(WDS 13473+1727), with an M2 companion. Apparently, the period is
about two thousand years (Hale 1994) and the orbit is very
eccentric. The separation between the two stars has been measured in
1991 to be 3.4 arc-sec (Mason et al.\ 2001), which translates to about
50 AU projected distance for a parallax of 64 $mas$ (ESA
1997). Planets in binary systems might go through different orbital
evolution, and therefore might be considered as special cases. Thus,
the low frequency of planets with large masses and short periods seems
to be even more real than is seen from the figure.

Statistical assessment of the significance of the low frequency found
in this part of the parameter space is under way. Very simple-minded
calculations that ignore both the observational selection effects and
the binarity of the three stars indicate a significance at the
2--3$\sigma$ level.  Taking into account the selection effect and the
binarity of the three stars makes the significance of the low
frequency even higher.

The paucity of large-mass planets with short periods and consequently
small orbits might be another clue to the formation and orbital
evolution of the extrasolar planets. There are now two different
scenarios that account for the existence of giant planets in close-in
orbits. One of them, accepted by most of the astronomical community,
assumes the planets were formed out of a disc of gas and dust at a
distance of 5~AU or larger, and have migrated through interaction with
the disc to their present position (e.g., Lin, Bodenheimer, \&
Richardson 1996). The other one is that the planets were formed by
some {\it in situ} disc instability (Boss 1997). In principle, our
findings can be accounted for by both scenarios.

From the migration point of view, our findings might indicate that
most large-mass planets halted their migration at orbital radius of
the order of 0.2 AU. Obviously, the more massive the planet is, the
more angular momentum and energy have to be removed from its orbital
motion to enable the migration. Angular momentum and energy could be
absorbed by the disc of gas and dust through generation of density
waves (e.g., Goldreich \& Tremaine 1980; Ward 1997) or by a
planetesimal disc through gravitational interaction with the planet
(e.g., Murray et al. 1998; Del Popolo, Gambera, \& Ercan 2001). A too
massive planet might move in until the local inner disc cannot absorb
its angular momentum and energy. Such a consideration might account
for a continuous dependence of the final orbital period on the
planetary mass.

Interestingly enough, some studies suggested different migration
scenarios for planets with small and large masses (Ward 1997).
Massive planets open a gap in the disc, and subsequently go through
slow, type II, migration, while small planets do not open a gap in the
disc and therefore go through a relatively fast, type I,
migration. The apparent paucity of short-period massive planets is
consistent with such an evolutionary separation between large and
small planets, if we can assume that the separation between the two
types of migration occurs at a mass of about 1 \MJ, and that type II
migration could halt at about 0.2 AU (e.g., Lin et al. 2000).

According to the instability scenario, the mass of the formed planet
depends on the available mass in the disc at the region of instability
(e.g., Boss 2000). At small distances the available mass might be
smaller, a fact that could result in low frequency of massive planets
with short periods.

The fact that all three planets with relatively large masses and short
periods are found in binary systems is intriguing. The interaction of
the secondary with the protoplanetary disc could modify the structure
and evolution of the disc, and therefore the formation and evolution
of the planet. We obviously need more data to see whether this feature
is statistically significant.

In all the aforementioned scenarios, the paucity of massive planets
with short-period orbits is a natural consequence of the formation and
evolutionary mechanism. However, detailed theoretical models have to
be worked out so we can compare the theory with the observations. If
confirmed by the discovery of more planets, the interesting input of
the present analysis is the actual boundaries of the low-frequency
part of the diagram.  A borderline at about 1.5 \MJ\ and at about 30
days can help us quantitatively understand the formation and
evolutionary process of extrasolar planets.

\section{Summary}

The logarithmic mass distribution derived here shows that the planet
candidates are indeed a separate population, probably formed in a
different way than the secondaries in spectroscopic
binaries. Surprisingly the eccentricity and period distributions, with
some restrictions, are very much the same.

Furthermore, the two period distributions follow strictly a straight
line. This indicates flat density distributions on a logarithmic scale,
inconsistent with the Duquennoy \& Mayor (1991) log-normal
distribution. Interestingly, flat logarithmic distribution 
is the only scale-free distribution, and could be argued to be the
most simple distribution. Maybe the two populations were formed by
two different mechanisms that still have this scale-free feature in
common (Heacox 1999).  

The eccentricity distribution of the sample of giant planets and that
of stellar companions are similar (Stepinski \& Black 2001c).  In
spite of the similarity, they are not identical, especially if
compared to the remarkable similarity between the two period
distributions.  The eccentricity distributions can be attained by
Keplerian orbits whose velocities are normally disturbed in the
tangential and the radial directions.

We found a significant paucity of large planets with short orbital
periods, and point out how, in principle, one can account for this
paucity.

\section*{Acknowledgments}
We acknowledge support from the Israeli Science Foundation through
grant no.\ 40/00.  This research has made use of the {\footnotesize
SIMBAD} database, operated at {\footnotesize CDS}, Strasbourg, France,
and the Washington Double Star Catalog maintained at the U.S. Naval
Observatory.

\vspace{0.7cm}
\noindent
{\large{\bf References}}
{\small

\bref
Basri, G., \& Marcy, G.~W. 1997, in AIP Conf. Proc 393,
Star Formation, Near and Far, eds. S. Holt \& L.G. Mundy (New York: AIP),
228

\bref
 Black, D. C. 1995, ARA\&A, 33, 359

\bref
Boss, A. P. 1997, Science, 276, 1836

\bref
Boss, A. P. 2000, ApJL, 536, L101

\bref
 Brown, T. M., Charbonneau, D., Gilliland, R. L., Noyes,
R. W., \& Burrows, A. 2001, ApJ, 552, 699

\bref
 Carney, B.~W., \& Latham, D. W. 1987, AJ, 92, 116

\bref
 Charbonneau, D., Brown, T.~M., Latham, D.~W., \& Mayor, M. 2000,
 ApJ, 529, L45

\bref
Del Popolo, A., Gambera, M., \& Ercan, N. 2001, MNRAS, 325, 1402

\bref
 Duquennoy, A., \& Mayor, M. 1991, A\&A, 248, 485

\bref
Els, S. G., Sterzik, M. F., Marchis, F., Pantin, E., Endl, M., 
\& K\"urster, M. 2001, A\&A, 370, L1

\bref 
 ESA 1997, The Hipparcos and Tycho Catalogues, ESA SP-1200

\bref
Gatewood, G., Han, I., \& Black, D. 2001, ApJ, 548, L61

\bref
 Goldberg, D. 2000, Ph.D. thesis, Tel Aviv University

\bref
 Goldberg, D., Mazeh, T., Latham, D.~W., Stefanik, R.~P., Carney,
B.~W., \& Laird, J.~B. 2001, submitted to A\&A

\bref
Goldreich, P., \& Tremaine, S. 1980, ApJ, 241, 425

\bref
 Good, P. 1994, Permutation Tests --- A Practical Guide to Resampling 
Methods for Testing Hypotheses, (New York: Springer-Verlag)

\bref
 Halbwachs, J.-L., Arenou, F., Mayor, M., Udry, S., \& Queloz,
D. 2000, A\&A, 355, 581

\bref
Hale, A. 1994, AJ, 107, 306

\bref
Han, I., Black, D., \& Gatewood, G. 2001, ApJ, 548, L57

\bref
 Heacox, W.~D. 1995, AJ, 109, 2670

\bref
 Heacox, W.~D. 1999, ApJ, 526, 928

\bref
Henry, G. W., Marcy, G. W., Butler, R. P., \& Vogt, S. S. 2000,
ApJ, 529, L41

\bref
 Jorissen, A., Mayor, M., \& Udry, S. 2001, A\&A, in press,
 astro-ph/0105301

\bref
 Kendall, M. G., \& Stuart, A. 1966, The Advanced Theory of
 Statistics, vol. 3, (London: Griffin)

\bref
 Latham, D.~W. 1985, in IAU Colloq. 88, Stellar Radial Velocities,
eds. A.~G.~D.~Philip \& D.~W.~Latham (Schenectady, L.~Davis Press) 21

\bref
 Latham, D.~W., Stefanik, R.~P., Torres, G., Davis, R.~J., Mazeh, T.,
Carney, B.~W., Laird, J.~B., \& Morse, J.~A. 2001, submitted to A\&A

\bref
Lin, D. N. C., Bodenheimer, P., \& Richardson, D. C. 1996, Nature,
380, 606

\bref
Lin, D. N. C., Papaloizou, J. C. B., Terquem, C., Bryden, G., \& Ida,
S. 2000, in Protostars and Planets IV 
eds.  V.~Mannings, A.~P.~Boss, S.~S.~Russell
(Tucson: University of Arizona Press), 1111

\bref
Lissauer, J.J. 1993, ARA\&A, 31, 129

\bref
Marcy, G.~W., \& Butler, R.~P. 1998, ARA\&A, 36, 57

\bref
Marcy, G.~W., \& Butler, R.~P. 2000, PASP, 112, 137

\bref
 Marcy, G. W., Cochran, W. D., \& Mayor, M. 2000 in Protostars and
Planets IV eds. V.~Mannings, A.~P.~Boss, S.~S.~Russell (Tucson:
University of Arizona Press), 1285

\bref 
Mason, B. D., Wycoff, G. L., Hartkopg, W. I., Douglass, G. G., \&
Worley, C. E. 2001, Washington Double Star Catalog 2001.0, 
U.S. Naval Observatory, Washington

\bref
Mayor, M.,  Queloz, D., \& Udry, S. 1998, in Brown Dwarfs
and Extrasolar Planets, eds. R. Rebolo, E.L. Martin, \& 
M.R. Zapatero-Osorio (San Francisco: ASP), 140


\bref
 Mayor, T., Udry, S., \& Queloz, D. 1998, in ASP Conf. Ser. 154, 
Tenth Cambridge Workshop on Cool Stars, Stellar Systems, and the Sun, 
eds. R. Donahue \& J. Bookbinder (San Francisco: ASP), 77

\bref
 Mazeh, T. 1999a, Physics Reports, 311, 317

\bref
 Mazeh, T. 1999b,, in ASP Conf. Ser. 185, IAU Coll. 170, Precise
Stellar Radial Velocities, eds. J.~B. Hearnshaw \& C.~D. Scarfe,  (San
Francisco: ASP), 131

\bref
 Mazeh, T. et al. 2000, ApJ, 532, L55

\bref
Mazeh, T., \& Goldberg, D. 1992, ApJ, 394, 592

\bref 
Mazeh, T., Goldberg, D., \& Latham, D. W. 1998, ApJ, 501, L199

\bref
 Mazeh, T., Mayor, M., \& Latham D.~W. 1996, ApJ, 478, 367

\bref
Mazeh, T., \& Zucker, S. 2001, in IAU Symp. 200, Birth and Evolution of Binary 
Stars, eds. B. Reipurth and H. Zinnecker (San Francisco: ASP), 519

\bref
 Mazeh, T., Zucker, S., Dalla Torre, A., \& van Leeuwen, F. 1999,
  ApJ, 522, L149

\bref
Murray, N., Hansen, B., Holman, M., \& Tremaine, S. 1998, Science,
279, 69

\bref
Pourbaix, D. 2001, A\&A, 369, L22

\bref
 Pourbaix, D., \& Arenou, F. 2001, A\&A, 372, 935 

\bref
 Schneider, J. 2001, in Extrasolar Planets Encyclopaedia 
http://www.obspm.fr/planets

\bref
 Stepinski, T. F., \& Black, D. C. 2001a, in IAU Symp. 200,
Birth and Evolution of Binary Stars, ed. B. Reipurth \& H. Zinnecker
(San Francisco: ASP) 167

\bref
 Stepinski, T. F., \& Black, D. C. 2001b, A\&A, 356, 903 

\bref
 Stepinski, T. F., \& Black, D. C. 2001c, A\&A, 371, 250

\bref 
Tokovinin, A. A. 1991, Sov. Astron. Lett., 17, 345 

\bref 
Tokovinin, A. A. 1992, A\&A, 256, 121

\bref
Ward, W.~R. 1997, Icarus, 126, 261

\bref
Zucker, S., \& Mazeh, T. 2000, ApJ,  531, L67 

\bref
 Zucker, S., \& Mazeh, T. 2001a, ApJ, in press (astro-ph/0107124)

\bref
 Zucker, S., \& Mazeh, T. 2001b, ApJ, in press (astro-ph/0106042)

}

\vfill

\end{document}